\documentstyle[epsfig,aps]{revtex} 

\def\be{\begin{equation}} 
\def\ee{\end{equation}} 
\def\ba{\begin{eqnarray}} 
\def\ea{\end{eqnarray}}

\begin{document} 
 
\title{Relativistic effects and quasipotential equations}
\author{G.\ Ramalho$^1$, A.\ Arriaga$^{1,2}$ and M.\ T.\ Pe\~na$^{3,4}$ \\
\vskip 2mm
{\small $^1$Centro de F{\'\i}sica Nuclear da Universidade de Lisboa, 
1649-003 Lisboa Codex, Portugal \\
        $^2$Departamento de F{\'\i}sica, 
Faculdade de Ci\^encias da Universidade de Lisboa, 
1749-016 Lisboa, Portugal \\
        $^3$Departamento de F{\'\i}sica, Instituto Superior T\'ecnico, 
1049-001 Lisboa, Portugal \\
        $^4$Centro de F{\'\i}sica das Interac\c c\~oes Fundamentais,
IST, 1049-001 Lisboa, Portugal.} \\[2mm]}
\date{October 24, 2001}
\maketitle

\begin{abstract} 
We compare the scattering amplitude  
resulting from the several quasipotential equations for scalar particles.  
We consider the Blankenbecler-Sugar, Spectator, Thompson,  
Erkelenz-Holinde and Equal-Time equations, which were solved numerically
without decomposition into partial waves.  
We analyze both negative-energy state components of the propagators
and retardation effects. 
We found that the scattering solutions of the Spectator and
the Equal-Time equations are very close to the nonrelativistic solution
even at high energies.
The overall relativistic effect increases 
with the energy. 
The width of the band for the relative uncertainty in the real part of the 
scattering $T$ matrix, due to different dynamical 
equations, is largest for backward-scattering angles 
where it can be as large as 40\%. 
\end{abstract} 
 
\section{Introduction} 
\label{intro} 
 
Although hadrons are  
small bags of confined quarks, at low and intermediate energies 
what is observed  
are mesons and baryons, suggesting that
the genuine QCD degrees of freedom may not be 
adequate for the study of nuclear structure at 
this energy regime. At low energies the nuclear dynamics is described 
in terms of nucleons
interacting via potentials, within the framework of 
the Schr\"odinger or Lippmann-Schwinger
equations. At intermediate energies, however, the
Non Relativistic (NR) quantum mechanical approach may fail. 
In particular, for processes
with high momentum and energy transfer, we may expect the 
NR scattering equation  
to be inadequate to describe the nucleon-nucleon (NN) scattering.  

The NR scattering equation, or  Lippmann-Schwinger (LS) equation,  
gives the NN scattering amplitude $T$, as: 
\be 
T({\bf p^ \prime},{\bf p};W)=V({\bf p^ \prime},{\bf p};W)- 
\int \frac{{\bf d^3 k}}{(2\pi)^3}  
V({\bf p^\prime},{\bf k};W)  
\frac{m}{{\bf k^2}-{\bf p^2} -i \varepsilon} 
T({\bf k},{\bf p};W), 
\label{LS} 
\ee 
where $m$ is the nucleon mass, ${\bf p}$, ${\bf p^{\prime}}$ and ${\bf k}$ are
respectively the initial, final and intermediate relative 3-momenta;
$W$ is the total energy which depends on ${\bf p}^2$, and
$V$ is the interaction kernel. This kernel, in a NR framework, describes the 
instantaneous interaction or potential. The homogeneous term 
involves also an effective
2-particle scalar propagator.
 
For a relativistic formulation, we have several alternative equations since
there is not a unique description dictated from first principles 
\cite{BrownJackson}.
An implementation of relativity which satisfies covariance explicitly 
leads to
the Bethe-Salpeter (BS) scattering equation \cite{BS51}: 
\be 
T(p^ \prime,p;P)=V(p^ \prime,p;P)+ 
i \int \frac{d^4 k}{(2\pi)^4}  
V(p^\prime,k;P) G(k;P) 
T(k,p;P). 
\label{BS} 
\ee 
Here $p$, $p^\prime$ and $k$ are respectively the final, initial and
intermediate relative 4-momenta, and $P$ the total system 4-momentum.  
As for $V$, the interaction kernel, it consists of the sum of all
irreducible interaction processes derived from a considered Lagrangian.
In the scalar case
the 2-particle propagator $G$ is given by
\be 
G(k;P)=\frac{1}{\left[m^2-(P/2+k)^2-i\varepsilon\right] 
\left[m^2-(P/2-k)^2-i\varepsilon\right]}. 
\label{Gscalar} 
\ee

Contrarily to the LS equation, which is 3-dimensional and 
includes only instantaneous interactions, the BS is a 
4-dimensional integral equation and its kernel 
includes retardation and a dependence on the virtual 
intermediate states energy (off-mass-shell effects).
Since the exact solution of the BS equation 
with the complete kernel is still out of reach of present calculation
capabilities, several approximations have been developed. 
Nevertheless, all these different
methods have in common the truncation of the kernel, which keeps only 
a restrict
set of meson exchange diagrams.
In Ref.\ \cite{TjonLadder}, the 4-dimensional 
BS equations was applied by Tjon and 
collaborators to the NN problem 
with a kernel derived from One-Boson-Exchange (OBE) 
generating the so called Ladder Approximation.
It has been proved, however, that the Ladder Approximation does not  
verify the one-body limit \cite{Gross69}. 
The addition to the OBE kernel of
a crossed-box diagram, implementing a particular off-mass shell 
extrapolation, has been done by Theu\ss l and Desplanques 
\cite{Theu00} for a scalar bound state problem.
Alternatively, for the same problem Phillips and Wallace
\cite{Phillips96}
simulated the cross-box diagram through a specific 
modification of the 2-particle propagator and the result 
is known as the so called 4D Equal-Time equation.
Calculations considering ladder and crossed-ladder
diagrams have been performed by  Nieuwenhuis and Tjon \cite{Nieu96},  
based on a Feynman-Schwinger formalism and Monte-Carlo  
techniques. The calculations are, however, still restricted 
to scalar particles and the
bound state cases.

Even with a truncated kernel the BS equation is, beyond Ladder Approximation, 
very hard to solve. This difficulty has triggered the development 
of covariant  
3-dimensional reduction equations. These can be classified into two classes: 
the QuasiPotential (QP) type equations and the Klein type equations.

In the QP equations the  3-dimensional reduction is achieved by fixing the
relative energy $k_0$ in a covariant way. These equations can formally
be written as a BS equation, if we replace $G$ by another propagator  
$g$ including a covariant energy constraint. 
In the last years two QP equations have been applied 
extensively to hadronic physics problems: the Blankenbecler and 
Sugar (BbS) 
\cite{Tjon80,Machleidt89,Robilotta94} 
and the Spectator (Sp), or Gross equations
\cite{Gross92,Stadler97}.

The BbS equation \cite{BbS66},
is constructed assuming that, in all intermediate states, 
the two particles are equally off-mass-shell.
The Sp equation \cite{Gross69}, is obtained restricting one 
particle to its positive-energy on-mass-shell state in all the 
intermediate states. Although this equation is exact for scalar 
particles in the one-body limit, when applied to the NN interaction 
it implies approximations. 
Furthermore, to account for the identity of the two particles, a 
symmetrized propagator
has to be used \cite{Gross92}.
In the BbS equation the interaction is instantaneous because the exchange boson
do not carry any energy, since both nucleons are equally off-mass-shell,
whereas in the Sp equation energy exchange is
allowed and therefore retardation is included.
Another QP equation, the 3D Equal-Time (ET) equation,
was proposed by Mandelzweig and Wallace to deal with the QED problem  
of spin $1/2$ particles \cite{MW87}, and with spin zero and spin $1/2$
particles with boson exchange cases \cite{MW89}. In this equation the
interaction is also instantaneous and includes the
crossed-box diagram in the eikonal approximation, or high energy limit 
\cite{Phillips96,MW87,MW89,Phillips98}, 
where the forward scattering components are dominant.
This 3D ET equation has been applied by 
by Tjon and collaborators to the NN system with an One-Boson-Exchange
kernel \cite{Hummel94}. Calculations have shown that the 3D ET equation
 results are the closest to the BS with OBE and crossed-box kernel
for the bound state of scalar particles \cite{Nieu96}.
Henceforth we refer the 3D ET equation as ET equation. 

The Klein equation, obtained also from the BS equation, 
is related with time ordered perturbation theory without 
anti-particles \cite{Phillips96,Phillips98}, 
and was used by the Nijmegen \cite{Rijken91,Rijken96} and Bonn 
\cite{Machleidt89} groups in their applications to two-meson-exchange  
calculations.

In this work we study relativistic effects included in 
the most well known QP equations -  BbS, Sp, Thompson (Th) \cite{Thompson70}
and Erkelenz-Holinde (EH) \cite{EH72} - and compare their results 
with ones obtained
with the NR Lippman-Schwinger equation. 
Since the exact solution of the relativistic equation is not known, there
is no exact relativistic answer to be taken as a reference. Therefore,
we focus on the analysis of the relativistic content of the different QP
equations. Our natural reference, consequently, is the nonrelativistic LS equation.
We emphasize, at this point, that the BS equation in ladder approximation,
which has been taken as a reference in the past \cite{WJ73}, has been 
shown to be inadequate
for that purpose because, on the one hand, does not reproduce 
the one-body limit, as 
mentioned before, and,
on the other hand, excludes non negligible 
contributions of the crossed-box diagrams,
as found by the authors of Ref. \cite{Nieu96}. 

As a first step, we consider in all our calculations 
(corresponding to different equations), 
only scalar particles interacting
through the Malfliet-Tjon potential 
\cite{MalflietTjon}. 
Only by using the same interaction in all the dynamical cases 
we are able to withdraw conclusions about the way 
the different formalisms introduce relativistic effects.
The relativistic effects considered come from kinematics, 
retardation and negative-energy state 
components of the propagators. We point out that retardation
is automatically excluded from the instantaneous BbS equation.
As for the negative-energy state effects, the different equations account 
for them in 
different ways, as it will be shown.

In Sec.\ \ref{secQP} we present the QP formalism, and give  
the explicit expression of all the propagators.
In Sec.\ \ref{secDecomp}, we use the method of Ref.\ \cite{WJ73}, 
reviewed in references \cite{Stadler99},
to distinguish between the equations that include retardation
from the others where the interaction is instantaneous. We also introduce
the representation of Wallace and Madelzweig 
\cite{Phillips96,MW89,Phillips98,Lahiff97} 
to separate the positive and negative-energy state effects, 
and discuss the suppression of the negative-energy states contributions.
In reference \cite{WJ73} a similar derivation was presented, but none of
them discuss explicitly retardation effects, which is one of the main goals
of the present work.
In Sec.\  \ref{secNPWD} we explain the numerical method used to evaluate
the $T$ scattering amplitudes,  which does not 
involve a partial wave decomposition.
This method is based on the work of Elster and
collaborators \cite{Elster98} in a NR framework. 
In Sec.\ \ref{secResults} we show and discuss the results for the scattering
amplitude at two different energies. As an example we also show the phase
shifts for the s-wave. In Sec.\ \ref{secCon} we summarize
and draw some final conclusions and remarks.

\section{Quasipotential equations} 
\label{secQP} 

The BS equation (\ref{BS}), can be written in a shorthand notation as
\be
T=V+VGT.
\label{compactBS}
\ee

A common approximation consists on restricting $V$ to the One Boson 
Exchange (OBE) diagrams, and is usually called the Ladder Approximation.
There are two main objections to this approximation: firstly, 
as mentioned before,
the one body limit is not recovered when one 
of the masses goes to infinity \cite{Gross69}; 
secondly, it has been shown that, in general, crossed-ladder diagrams may have 
important contributions \cite{Gross69} and hence should not be 
neglected.

Alternative ways consists on the replacement of BS Eq.\ (\ref{compactBS}) 
by the set of equations:
\ba
T&=&K+KgT \label{QP1} \\
K&=&V+V(G-g)K, \label{QP2}
\ea
which is perfectly equivalent to the BS equation, and where $g$ is a different
2-particle propagator, containing the 3-dimensional reduction.
Since there are no first principle rules to dictate how this 
reduction should be performed, a large ambiguity gives room to many options. 
Usually, different approximations compromise different appropriate choices 
of the constraint on the energy-component of the 4-momentum in the 
propagator $g$, and different truncations of the kernel $K$.

The rate of convergence of the new kernel $K$ is determined 
by the difference $G-g$. If $g$ is a good approximation to $G$,
then only the first term in Eq.\ (\ref{QP2}) can be kept and Eqs.\ (\ref{QP1})
and (\ref{QP2}) reduces to one simpler 
3-dimensional, but still covariant, equation. This is the 
standard procedure in any application of QP equations. 
If needed and possible, the lowest order truncation of the kernel 
$K=V$ can be corrected by higher order terms.

Any finite order truncation of the kernel beyond the OBE approximation 
generates unphysical singularities in the scattering equation, due to the delta
function constraint on the energy\cite{Phillips96}. 
For this reason 3-dimensional QP equations with kernels not restricted to
the OBE form became controversial \cite{Phillips96,Phillips98,Lahiff97}.  
However, we point out that it has been shown that the contribution of 
the two-pion subtracted \cite{Ramalho99}
and crossed-box diagrams can be simulated by effective $\omega$ 
and $\rho$ vector
meson exchanges \cite{Ramalho99}.
This result supports the 
usefulness of the OBE
approximation.

In the present work we apply several QP equations,
with an OBE kernel, to the scattering of scalar particles. More specifically,
we considered the scalar Malfliet-Tjon potential and
took the form of this potential as the kernel of all the equations considered. 

We perform the calculations in the c.m.\ reference frame, 
where the expressions are
simpler. In this frame the Bethe-Salpeter to 2-particle propagator is 
written as  
\be 
G(k;P)=\frac{(2m)^2} 
{\left[E_{\footnotesize \mbox{k}}^2-(W/2+k_0)^2-i \varepsilon \right] 
\left[E_{\footnotesize \mbox{k}}^2-(W/2-k_0)^2-i \varepsilon \right]}. 
\label{gBS} 
\ee 
Here $\mbox{k}$ represents the magnitude of ${\bf k}$,   
$E_{\footnotesize \mbox{k}}=\sqrt{m^2+{\bf k}^2}$ and 
$W=\sqrt{P^2}=2 E_{\footnotesize \mbox{p}}$, 
if both initial particles
are on-mass-shell.  
This expression differs from the original of Eq.\ (\ref{Gscalar}) 
by the factor  
$(2m)^2$, which is introduced for convenience. In fact, in the scalar case
both the kernel and the amplitude $T$ are dimensionless, 
whereas the amplitude generated by
the LS equation, which refers to fermions, 
has dimensions of inverse of energy squared.
In order to compare the solutions of the two equations, we must redefine $V$
and $T$ of the scalar equation by dividing 
them by $(2m)^2$, meaning that we have to
redefine the propagator by multiplying it by the same factor.
This procedure implies the use of dimensionless couplings for 
both the scalar and
Dirac cases.

As mentioned in the introduction, in the present work 
we consider the following QP
equations: the Sp, Th, BbS, EH and the ET.  
All these equations can be obtained replacing   
the propagator $G$ of the Eq.\ (\ref{BS}) by the  
following $g$ functions: 
\ba 
g_{Th}(k;W)&=&i 2 \pi \frac{m}{E_{\footnotesize \mbox{k}}} \frac{m}{W} 
\frac{\delta(k_0)}{E_{\footnotesize \mbox{k}}-\frac{W}{2}-i \varepsilon}  
\label{gTh}\\ 
g_{BbS}(k;W)&=&i 2 \pi \frac{m}{E_{\footnotesize \mbox{k}}}  
\frac{m \delta(k_0)}{E_{\footnotesize \mbox{k}}^2-\frac{W^2}{4}-i \varepsilon} 
\label{gBbS}
\\ 
g_{EH}(k;W)&=&i 2 \pi \frac{m}{E_{\footnotesize \mbox{k}}}  
\frac{m \delta(k_0-E_{\footnotesize \mbox{k}}+W/2)}{E_{\footnotesize \mbox{k}}^2-\frac{W^2}{4}-i \varepsilon}  
\label{gEH}\\ 
g_{Sp}(k;W)&=& i 2 \pi \frac{m}{E_{\footnotesize \mbox{k}}} \frac{m}{W} 
\frac{\delta(k_0-E_{\footnotesize \mbox{k}}+W/2)}{E_{\footnotesize \mbox{k}}-\frac{W}{2}-i \varepsilon}.  
\label{gGross} 
\ea 

The BbS and EH propagators are identical except for the argument of the delta
function containing the energy constraint. The same can be said about the
Th and the Sp propagators.
We also consider the ET scalar equation 
\cite{MW89}, corresponding to the propagator 
\be 
g_{ET}(k;W)=i 2 \pi \frac{m}{E_{\footnotesize \mbox{k}}}  
\frac{m \delta(k_0)}{E_{\footnotesize \mbox{k}}^2-\frac{W^2}{4}-i \varepsilon}  
\left(2- \frac{W^2}{4E_{\footnotesize \mbox{k}}^2} \right), 
\label{gET} 
\ee 
that differs from BbS propagator (\ref{gBbS}) by a kinematical factor,
which becomes one when the particles are on-mass-shell ($W=2 E_{\footnotesize \mbox{k}}$).

In all cases the energy-fixing condition is included through a  
$\delta$-function. We note here that even the NR equation (\ref{LS})
can be obtained from the BS Eq.\ (\ref{BS}) by replacing  
the 2-particle $G$ propagator by the NR propagator   
\be 
g_{NR}(k;W)=i 2 \pi \frac{m \delta(k_0)}{{\bf k^2}-{\bf p^2}  
-i\varepsilon},
\label{gNR} 
\ee 
which can also be written as
\be 
g_{NR}(k;W)=i 2 \pi   
\frac{m \delta(k_0)}{E_{\footnotesize \mbox{k}}^2-\frac{W^2}{4}-i \varepsilon}, 
\label{gNR2} 
\ee 
but this does not mean that the LS equation is covariant.  

All the propagators defined above are for scalar particles.
The first five propagators can be organized into two classes, by means 
of a function $f(\mbox{k};W)$, the BbS and Sp types \cite{BrownJackson,WJ73,Klein74}.
The first class includes instantaneous equations
while the second one includes equations with retardation.
\ba
g_{BbS}(k;W) &\to &i 2 \pi \frac{m}{E_{\footnotesize \mbox{k}}}  
\frac{m f(\mbox{k};W)}{E_{\footnotesize \mbox{k}}^2-\frac{W^2}{4}-i \varepsilon} \delta(k_0) 
\label{BbStype}
\\
g_{Sp}(k;W) &\to &i 2 \pi \frac{m}{E_{\footnotesize \mbox{k}}}   
\frac{m f(\mbox{k};W)}{E_{\footnotesize \mbox{k}}^2-\frac{W^2}{4}-i \varepsilon}
\delta(k_0-E_{\footnotesize \mbox{k}}+W/2).
\label{Grosstype}
\ea
Any choice of $f$ is possible providing that 
\be
f(\mbox{p};W)=1,
\label{onshell}
\ee
which results from the simultaneous constraint 
of the 2 nucleons in its physical state  
(positive-energy on-mass-shell state).   
Taking $f(\mbox{k};W) \equiv 1$ in Eqs.\ (\ref{BbStype}) 
and (\ref{Grosstype}),
we obtain respectively the original BbS and EH propagators.
Taking 
\be
f(\mbox{k};W)=\frac{W+2E_{\footnotesize \mbox{k}}}{2W},
\label{ThompFactor}
\ee
in Eqs.\ (\ref{BbStype}) and (\ref{Grosstype}), we get respectively the Th
and Sp propagators. Finally the ET propagator is obtained from 
Eq.\ (\ref{BbStype})
using the following $f(\mbox{k};W)$ function:
\be
f(\mbox{k};W)= \left(2-\frac{W^2}{4E_{\footnotesize \mbox{k}}^2} \right).
\ee

We point out that the function $f(\mbox{k};W)$, 
to a great extent arbitrary, is related with the 
weight of the contribution of the negative-energy state components, 
as we shall see.  

For numerical applications we rewrite the propagators 
(\ref{gTh})-(\ref{gGross}) and (\ref{gET}) in a different 
form. Firstly, we represent by $\bar g$ the propagator 
without the $\delta$-function factor.  Secondly, we define the nonsingular
function $\bar f(\mbox{k};W)$ accordingly to
\be
\bar g(\mbox{k};W)= \frac{\bar f(\mbox{k};W)}{E_{\footnotesize \mbox{k}}-\frac{W}{2}-i\varepsilon}.
\label{gbar}
\ee
and $\bar f(\mbox{k};W)$ is expressed in terms of $f(\mbox{k};W)$ by 
\be
\bar f(\mbox{k};W)= \frac{m}{E_{\footnotesize \mbox{k}}}\frac{mf(\mbox{k};W)}{E_{\footnotesize \mbox{k}}+\frac{W}{2}}.
\ee
These $\bar f(\mbox{k};W)$ 
functions verify the
relativistic elastic cut condition, expressed in Eq.\ 
(\ref {onshell}),
corresponding to  
\be 
\bar f(\mbox{p};W)=2  
\left( 
\frac{m}{W} 
\right)^2.
\label{eqf2}
\ee

\section{Negative-energy components of the propagators and
 retardation effects in QP equations } 
\label{secDecomp} 
 
In order to clarify the interpretation of the 
positive and negative-energy state effects we follow
here the analysis of Ref.\ \cite{Stadler99}.  
We start with the homogeneous term of the Eq.\ 
(\ref{BS}), in the c.m.\ frame where $P=(W,{\bf 0})$
\be 
{\cal R}({\bf k};P)= 
\int \int {\bf d^3 k}  d k_0 V(p^\prime_0,{\bf p^\prime};  
k_0, {\bf k}; W) 
G(k;P) 
T(k_0, {\bf k};p_0, {\bf p}; W); 
\ee 
in the above equation we explicitly separate  
the energy and 3-momentum components of all momenta.  
Using the representation of Wallace and Mandelzweig   
\cite{Phillips96,MW89,Phillips98}, the propagator of 1-particle with 4-momentum $P/2+k$, 
can be written as 
\be
G_1(k;P)=
-\left\{  
\frac{N_1^+}{k_0-E_{\footnotesize \mbox{k}}+W/2+i\varepsilon}- 
\frac{N_1^-}{k_0+E_{\footnotesize \mbox{k}}+W/2-i\varepsilon} 
\right\}, 
\label{G1pm}
\ee
and the propagator of particle 2 with of momentum $P/2-k$ as 
\be
G_2(k;P)=
\left\{  
\frac{N_2^+}{k_0-E_{\footnotesize \mbox{k}}+W/2-i\varepsilon}- 
\frac{N_2^-}{k_0-E_{\footnotesize \mbox{k}}-W/2+i\varepsilon} 
\right\}. 
\label{G2pm}
\ee 
For scalar particles $N_i^+=N_i^-=m/E_{\footnotesize \mbox{k}}$ whereas for fermions
$N_i^+$ and $N_i^-$ are respectively the positive and 
negative-energy projectors \cite{Phillips96,MW89,Phillips98}.

Defining  
\ba 
\omega_{\footnotesize \mbox{k}}^+ 
&= & E_{\footnotesize \mbox{k}}-\frac{W}{2} - i \varepsilon \label{wp}\\ 
\omega_{\footnotesize \mbox{k}}^- &= & -\left(E_{\footnotesize \mbox{k}}+\frac{W}{2} \right) + i \varepsilon.  
\label{wm}  
\ea 
we can rewrite the BS 2-particle propagator $G$ as  
\ba 
G(k;P)=&-& 
\left\{ \frac{}{} \right. 
\underbrace{\frac{N_1^+}{k_0-\omega_{\footnotesize \mbox{k}}^+}}_{E_1>0}   
\underbrace{-\frac{N_1^-}{k_0-\omega_{\footnotesize \mbox{k}}^-}}_{E_1<0}  
\left.   \frac{}{}  \right\}  \nonumber \\ 
& \times & 
 \left\{  
\frac{}{} \right. 
\underbrace{\frac{N_2^+}{k_0+\omega_{\footnotesize \mbox{k}}^+}}_{E_2>0}  
\underbrace{-\frac{N_2^-}{k_0+\omega_{\footnotesize \mbox{k}}^-}}_{E_2<0} \left. 
\frac{}{} 
\right\}. 
\label{Gdecomp} 
\ea  
In the expression above we indicate, for the two terms of each 
particle propagator,
the sign of the particle energy at the corresponding pole. 

\begin{figure} 
\centering{
\epsfig{file=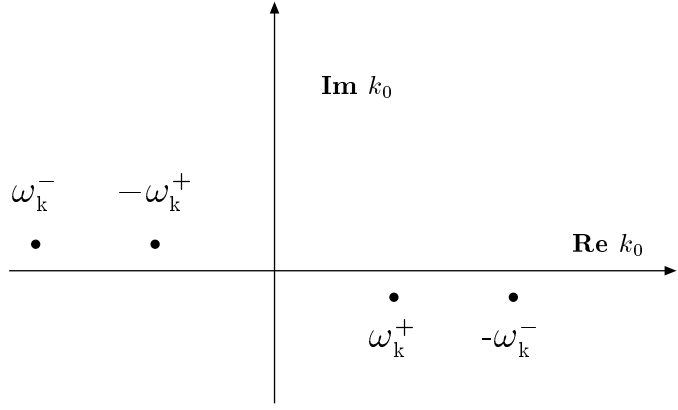}}
\caption{Poles structure of the $G$ propagator.} 
\label{polos} 
\end{figure}

Following \cite{WJ73,Stadler99}, the
QP equations can be obtained in a different way than the one 
discussed in the previous section,
namely by fixing the relative energy $k_0$  
in $V$ and, consequently in $T$. If we choose $k_0=0$ in $V$ and $T$, we 
can factorize 
the energy and 3-momentum integrations in the following form
\be 
{\cal R}({\bf k};P)= 
\int {\bf d^3 k} V(0,{\bf p^\prime};0, {\bf k}; W) 
T(0,{\bf k}; 0, {\bf p}; W) 
{\cal G}(\mbox{k};P),
\label{k0} 
\ee 
where ${\cal G}$ is  
\be 
{\cal G}(\mbox{k};P)= 
\int  d k_0 G(k;P) 
\label{Gdef} 
\ee 
With this choice the interaction between the nucleons  
is instantaneous. We call these equations the instantaneous-type equations. 
The integration in the $k_0$ variable can be done analytically   
using the residues theorem .  
The pole structure of the integrand function 
is presented in the Fig.\ \ref{polos}.  
Choosing the lower half contour for the $k_0$  
integration, we can conclude that  
\be 
{\cal G}(\mbox{k};P)=I_1(\omega_{\footnotesize \mbox{k}}^+)+
I_2(-\omega_{\footnotesize \mbox{k}}^-), 
\label{resGdef} 
\ee 
where the residues $I_1$ and $I_2$ are 
\ba 
I_1(\omega_{\footnotesize \mbox{k}}^+)&=& i 2 \pi 
\left\{ \frac{N_1^+  N_2^+}{2E_{\footnotesize \mbox{k}}-W -i \varepsilon}
+\frac{N_1^+ N_2^-}{W}  \right\} \label{I1}  \\  
I_2(-\omega_{\footnotesize \mbox{k}}^-)&=& i 2 \pi 
\left\{ 
-\frac{N_1^-   N_2^+}{W}+\frac{N_1^- N_2^-}{W+2 E_{\footnotesize \mbox{k}}}  \right\}.    
\label{I2} 
\ea 
   
By keeping both $I_1(\omega^+_{\footnotesize \mbox{k}})$ 
and $I_2(-\omega^-_{\footnotesize \mbox{k}})$ in Eq.\ 
(\ref{k0}) we obtain
the BbS equation. Neglecting the residue 
$I_2(-\omega^-_{\footnotesize \mbox{k}})$ we
get the Th equation.

The other class of equations can be obtained when we fix the  
$k_0$ in order to correspond to the positive-energy state  
of the particle 1.  
This corresponds to the choice $k_0=\omega_{\footnotesize \mbox{k}}^+$.
So, in that case   
\be 
{\cal R}({\bf k};P)= 
\int {\bf d^3 k}  
V(\omega_{p^\prime}^+,{\bf p^\prime};\omega_{\footnotesize \mbox{k}}^+,{\bf k}; W) 
T(\omega_{\footnotesize \mbox{k}}^+,{\bf k};\omega_{\footnotesize \mbox{p}}^+,{\bf p}; W) 
{\cal G}(\mbox{k};P). 
\label{GBS} 
\ee 
All equations in this class include retardation. 
because they includes the transfered energy term in the  
interaction kernel. If in Eq.\ (\ref{GBS}) we include the 
two residues $I_1(\omega_{\footnotesize \mbox{k}}^+)$ and
$I_2(-\omega_{\footnotesize \mbox{k}}^+)$  we obtain  
Erkelenz-Holinde (EH) equation, and  if we neglect the last one,  
we obtain the Sp equation. In that respect 
Sp equation is more consistent since it considers the same 
value of $k_0$ in all integrand functions.

We summarize in Table \ref{tabela} the comparison among the 
several QP propagators accordingly
to the the discussion above.

Using the decompositions (\ref{I1}) and (\ref{I2}) we may write 
\ba
{\bar g}_{BbS}(\mbox{k};W)&=& 
{\bar g}_{EH}(\mbox{k};W)=
i 2 \pi 
\left\{ 
\frac{N_1^+ N_2^+}{2E_{\footnotesize \mbox{k}}-W -i \varepsilon} +
\frac{N_1^+ N_2^-}{W}-
\frac{N_1^- N_2^+}{W}+
\frac{N_1^- N_2^-}{2E_{\footnotesize \mbox{k}}+W}  
\right\} \label{BbSdecomp}
\\
{\bar g}_{Sp}(\mbox{k};W)&=& 
{\bar g}_{Th}(\mbox{k};W)=
i 2 \pi 
\left\{
\frac{N_1^+ N_2^+}{2E_{\footnotesize \mbox{k}}-W -i \varepsilon} +
\frac{N_1^+ N_2^-}{W}  
\right\}.
\ea 
To avoid any possible confusion we call the reader's attention to the fact
that, for spin $1/2$ particles, 
some authors call BbS to a somewhat different 2-particle propagator 
\cite{BrownJackson,Phillips96,Tjon80,BbS66}.

For the BbS/EH equations we need to consider the sum  
of both terms in (\ref{resGdef}).  
It is interesting to note that, 
if the weight functions are equal, $N_i^{+}=N_i^{-}$, as it happens 
in the scalar case, the residue $I_1$ at the negative-energy pole of particle 2
cancels exactly against the residue $I_2$ at the positive-energy pole of the particle 1. 
As a result, we can write for the scalar case    
\be 
{\bar g}_{BbS}(\mbox{k};W) 
=i 2 \pi 
\left(\frac{m}{E_{\footnotesize \mbox{k}}}\right)^2  
\left\{ 
\frac{1}{2E_{\footnotesize \mbox{k}}-W -i \varepsilon} + 
\frac{\eta}{2E_{\footnotesize \mbox{k}}+W}  
\right\}. 
\label{BbSEH} 
\ee 
where $\eta=1$ corresponds to the maximal inclusion of negative-energy 
state term of
the propagators, and the case $\eta=0$ totally excludes them. In the same situation we
obtain for the Sp/Th equations the following 
\be  
{\bar g}_{Sp}(\mbox{k};W)=i 2 \pi  
\left( \frac{m}{E_{\footnotesize \mbox{k}}} \right)^2 
\left\{ 
\frac{1}{2E_{\footnotesize \mbox{k}}-W -i \varepsilon} + 
\frac{\eta}{W}  
\right\}. 
\label{Gross} 
\ee 
Comparing (\ref{BbSEH})  with (\ref{Gross})  in the case of  
negative-energy suppression ($\eta=0$)   
we conclude that all the propagators are equal. 

As for the ET propagator, it has been obtained in Refs.\  
\cite{MW87,MW89} and reads
\be
{\bar g}_{ET}(\mbox{k};W)= i 2 \pi 
\left\{
\frac{N_1^+ N_2^+}{2 E_{\footnotesize \mbox{k}}-W -i\varepsilon}+
\frac{N_1^+ N_2^-}{2E_{\footnotesize \mbox{k}}}+
\frac{N_1^- N_2^+}{2 E_{\footnotesize \mbox{k}}}+
\frac{N_1^- N_2^-}{2E_{\footnotesize \mbox{k}}+W} \right\}.
\ee
Compared to the BbS propagator, the ${\bar g}_{ET}$ contains two additional
contributions, which the authors justify as an effective inclusion
of the crossed-box diagram.
Restricting to the scalar case, this expression simplifies to 
\be
{\bar g}_{ET}(\mbox{k};W) =
i 2 \pi 
\left(\frac{m}{E_{\footnotesize \mbox{k}}}\right)^2 
\left\{ 
\frac{1}{2E_{\footnotesize \mbox{k}}-W-i \varepsilon}+
\frac{\eta}{E_{\footnotesize \mbox{k}}}  +
\frac{\eta}{2E_{\footnotesize \mbox{k}}+W}
\right\}.  
\ee  
If we restrict ourselves to the positive-energy components ($\eta=0$)  
then the ET propagator is  equivalent to all the other propagators. 
In particular, for a complete suppression of negative-energy states,  
BbS and ET equations generate the same results.

\begin{table} 
\begin{tabular}{|c|l|cc|} 
        & $V$, $T$ & $I_1(\omega_{\footnotesize \mbox{k}}^+)$ & 
$I_2(-\omega_{\footnotesize \mbox{k}}^-)$ \\ 
\hline 
Thompson & $k_0=0$      & $\sqrt{}$  & $\times$  \\ 
\hline 
Blankenbecler-Sugar      & $k_0=0$      & $\sqrt{}$  & $\sqrt{}$  \\ 
\hline  
Erkelenz-Holinde & $k_0=\omega_{\footnotesize \mbox{k}}^+$      & $\sqrt{}$  
& $\sqrt{}$  \\ 
\hline 
Spectator & $k_0=\omega_{\footnotesize \mbox{k}}^+$      & $\sqrt{}$  & 
$\times$  \\ 
\end{tabular} 
\caption{Comparison among QP equations. 
Here $k_0=\omega_{\footnotesize \mbox{k}}^+$ means that
retardation is taken into account, and  
$\times $ means suppression of the residue.} 
\label{tabela} 
\end{table}

\section{Solving The Scattering Equation Without Partial Wave Decomposition} 
\label{secNPWD} 
 
In the previous section we derived the QP equations in a way to 
make explicit, in each
of the considered equations, the contribution of the positive 
and negative-energy terms
of the BS propagator, as well as the effects of retardation 
present in the kernel. However,
as mentioned before, this method generates exactly the 
same QP equations as the one
described in Sec.\ \ref{secQP}, and we chose to proceed here 
with the formalism presented in the latter section.

Our starting point is, consequently, Eq.\ (\ref{QP1}) 
with $g$ given by one of the functions
of Eqs.\ (\ref{gTh})-(\ref{gET}), which after performing the 
$k_0$ integration can be written
as: 

\be 
T({\bf p^ \prime},{\bf p};W)=V({\bf p^ \prime},{\bf p};W)- 
\int \frac{\bf d^3 k}{(2\pi)^3}  
V({\bf p^\prime},{\bf k};W)  
\bar g(\mbox{k};W) 
T({\bf k},{\bf p};W). 
\label{QPeq} 
\ee  
Here $\bar g$ is given by Eq.\ (\ref {gbar}). 
The value of the relative energy variable,
which is not explicit in Eq.\ (\ref {QPeq}), is $k_0=0$ 
for the instantaneous-type equations, and  
$k_0=E_{\footnotesize \mbox{k}}-W/2$ for the retarded-type equations. 
 
The QP equations are 3-dimensional  
integral equations, corresponding to a radial  
integration and a 2-dimensional angular integration.
Since the physics is independent of the scattering plane, we may choose
the incoming 3-momentum ${\bf p}$ to have the z-direction and
the scattering plane as the $xz$ plane, meaning that the final momentum 
is only defined by its magnitude and by the angle $\theta$, 
which then gives the scattering angle.
If the interaction is scalar (spinless particles), 
the $T$-matrix and the kernel $V$ 
are independent of the angle $\varphi$ that specifies the scattering plane.  
     
Representing the magnitudes   
of the initial and final 3-momenta by p and 
$\mbox{p}^{\prime}$ respectively,
and defining $u=\cos \theta$, we can write then  
\ba 
T({\bf p^\prime},{\bf p};W)&=&T(\mbox{p}^\prime,u;\mbox{p},1;W), \\
V({\bf p^\prime},{\bf p};W)&=&V(\mbox{p}^\prime,u;\mbox{p},1;W).
\label{Tdef} 
\ea
The intermediate 3-momentum is characterized in terms  
of a polar angle, $v=\cos \theta_{\footnotesize \mbox{k}}$, and an azimuthal angle $\varphi_{\footnotesize \mbox{k}}$. 
With this notation we write   
\ba 
T({\bf k},{\bf p};W)&=&T(\mbox{k},v,\mbox{p},1;W), \\  
V({\bf p^\prime},{\bf k};W)&=&V(\mbox{p}^\prime,u;\mbox{k},v,\varphi_{\footnotesize \mbox{k}};W). 
\label{Vdef} 
\ea  
Using Eqs.\ (\ref{Tdef}) and (\ref{Vdef}) and 
factorizing the integration in $\varphi_{\footnotesize \mbox{k}}$  
we can rewrite (\ref{QPeq}) as 
\ba 
& &T(\mbox{p}^\prime,u;\mbox{p},1;W)=
V(\mbox{p}^\prime,u;\mbox{p},1;W)\nonumber \\ 
& &-\int_0^{\infty} \frac{\mbox{k}^2 \mbox{dk}}{(2\pi)^2}  
\int_{-1}^1 dv \; 
\bar V(\mbox{p}^\prime,u;\mbox{k},v;W) 
\frac{\bar f(\mbox{k};W)}{E_{\footnotesize \mbox{k}}-\frac{W}{2}-i\varepsilon} 
T(\mbox{k},v;\mbox{p},1;W)  
\ea  
where  
\be 
\bar V(\mbox{p}^\prime,u;\mbox{k},v;W)= 
\frac{1}{2\pi} \int_0^{2\pi} 
V(\mbox{p}^\prime,u;\mbox{k},v,\varphi_{\footnotesize \mbox{k}};W) d \varphi_{\footnotesize \mbox{k}}.   
\ee 

Treating the singularity of the integral function in the usual way, we write
\ba 
& & 
T(\mbox{p}^\prime,u;\mbox{p},1;W)=V(\mbox{p}^\prime,u;\mbox{p},1;W) 
\nonumber \\ 
& & 
-\int_{-1}^1 dv \; 
{\cal P} 
\int_0^\infty \frac{\mbox{k}^2 \mbox{dk}}{(2\pi)^2}  
\bar V(\mbox{p}^\prime,u;\mbox{k},v;W) 
\frac{\bar f(\mbox{k};W)}{E_{\footnotesize \mbox{k}}-\frac{W}{2}} 
T(\mbox{k},v;\mbox{p},1;W) \nonumber \\ 
& & 
-i \pi \frac{(2m)^2 \mbox{p}}{W}  
\int_{-1}^1  \frac{dv}{(2\pi)^2} \; 
\bar V(\mbox{p}^\prime,u;\mbox{p},v;W)  
T(\mbox{p},v;\mbox{p},1;W) 
\label{newQP} 
\ea 
where we used 
\be 
\left. 
\frac{\mbox{k}^2 \bar f(\mbox{k};P)}{\left| \frac{\mbox{d}\;\;}{\mbox{dk}} 
\left(E_{\footnotesize \mbox{k}}-\frac{W}{2}\right) \right|} 
\right|_{\mbox{k}=\mbox{p}} 
=\frac{(2m)^2 \mbox{p}}{W},
\ee 
accordingly to the eq.\ (\ref{eqf2}).

By standard discretization we transform the integral equation into an  
algebraic linear set of equations, as explained in the appendix 
\ref{NoPartialWave}, and obtain, for a fixed  
on-mass-shell initial state, the half-off-mass-shell scattering amplitude  
$T(\mbox{p}^{\prime}_i,u_j;\mbox{p},1;W)$.  
 
One of the numerical checks that has been performed is the verification
of the optical theorem. 
The number of grid points for the magnitudes of the 3-momenta 
and for the cosine
of the polar angles are fixed by imposing an
accuracy of the order of 1\% in the
verification of the optical theorem. 
Typical values used were 17-25 for the magnitude of the momenta and 20-50 
for the angular
discretization.
We point out however, that the convergence of the on-shell $T$ matrix does
not require so many grid points.
  
\section{Results and discussion} 
\label{secResults} 

We solve the scattering equation (\ref{BS}) with the propagators  
(\ref{gTh})-(\ref{gET}). For instantaneous-type equations we
considered the kernel to be the Malfliet-Tjon potential \cite{MalflietTjon}:

\be
V({\bf k},{\bf p};W)=\frac{V_R}{\mu_R^2+({\bf k-p})^2}-
\frac{V_A}{\mu_A^2+({\bf k-p})^2},
\label{mfp}
\ee
with the parameters of Ref.\ \cite{Elster98}:

\ba
V_R=3,1769 \;\;\; & & \mu_R=305,86 \;\; {\mbox MeV}\\
V_A=7,9210 \;\;\; & & \mu_A=613,69 \;\; {\mbox MeV}
\ea

The choice of the interaction is due, on one hand, to its simplicity, and, on the other,
to the fact that it as has been
used recently in the literature \cite{Elster98}, 
within a nonrelativistic framework and 
without a partial wave decomposition for the T-matrix. This way
we could compare and control our numerical results. 

In order to study retardation effects however, 
we have to consider a covariant version of
Eq.\ (\ref {mfp}), obtained by replacing the square of the 3-momentum by the
negative square of the 4-momentum (${\bf q}^2 \to -q^2$).  
Therefore we use for retarded-type QP equations
\be
V({\bf k},{\bf p};W)
=\frac{V_R}{\mu_R^2+({\bf k-p})^2-(E_{\footnotesize \mbox{p}}-E_{\footnotesize \mbox{k}})^2}-
\frac{V_A}{\mu_A^2+({\bf k-p})^2-(E_{\footnotesize \mbox{p}}-E_{\footnotesize \mbox{k}})^2}.
\ee

\begin{figure} 
\centerline{
\epsfig{file=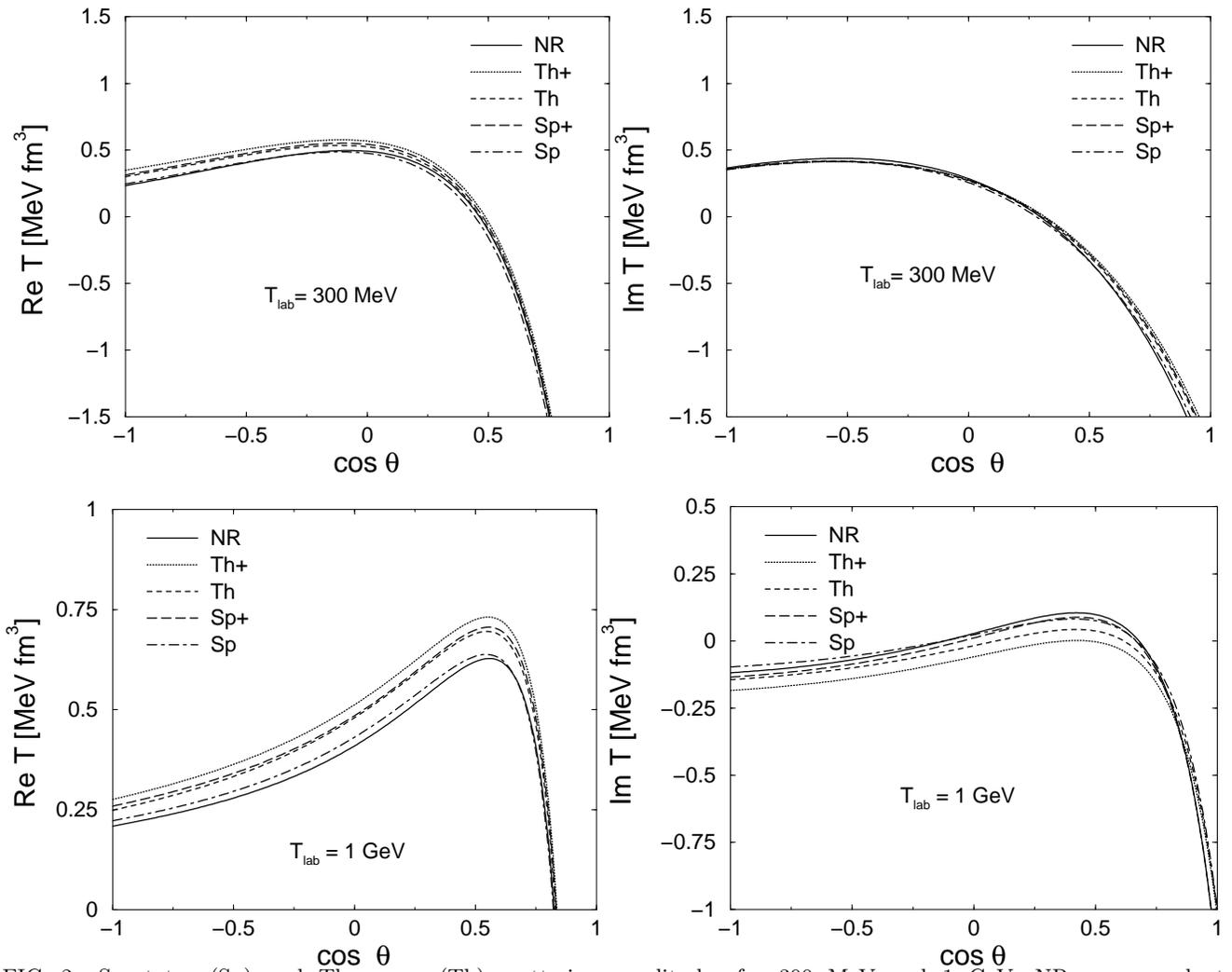}}
\caption{Spectator (Sp) and Thompson (Th) scattering amplitudes 
for 300 MeV and 1 GeV.  
NR corresponds to \mbox{Lippmann-Schwinger} amplitude.} 
\label{GrossTh} 
\end{figure}

\begin{figure} 
\centerline{ 
\epsfig{file=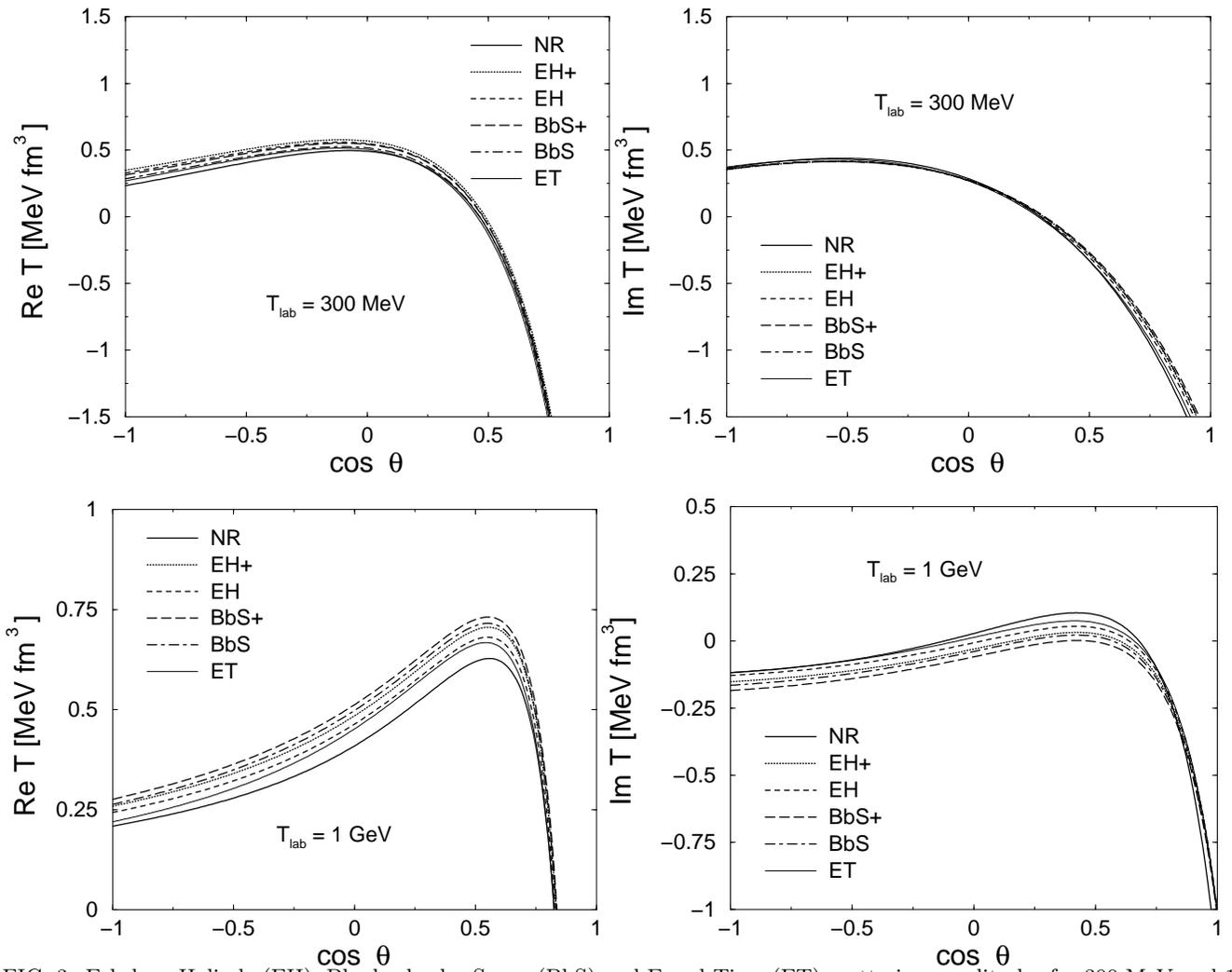}}  
\caption{Erkelenz-Holinde (EH), Blankenbecler-Sugar (BbS) 
and Equal-Time (ET)  scattering amplitudes for 300 MeV and 1 GeV.  
NR corresponds to Lippmann-Schwinger amplitude.} 
\label{EHBbS} 
\end{figure}

\begin{figure} 
\centerline{ 
\epsfig{file=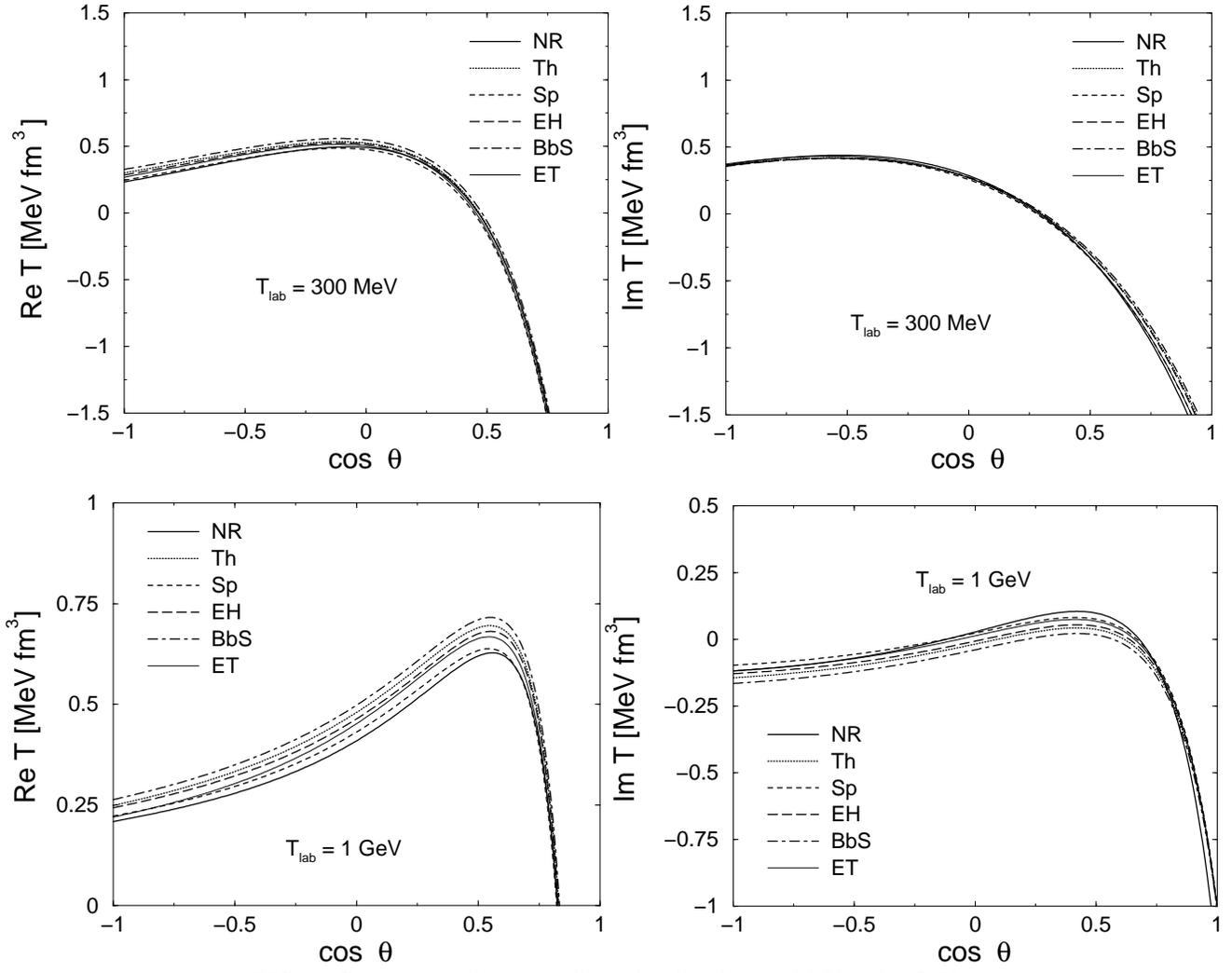}}  
\caption{Comparation between all amplitudes, 
for 300 MeV and 1 GeV.} 
\label{FigGeral} 
\end{figure}

To evaluate the scattering amplitude without partial wave  
decomposition we apply the method presented in the appendix  
\ref{NoPartialWave}. We consider throughout our calculations two energies:  
$T_{lab}=300$ MeV, where we expect small  
relativistic effects, and $T_{lab}=1$ GeV, which corresponds to
the intermediate energy range.  
The results are presented in the Figs.\ \ref{GrossTh}, 
\ref{EHBbS} and \ref{FigGeral}.
We will focus mainly on the real part of $T$, where 
the effects are more significant, but the imaginary part is also
presented for completeness. 

We organize our study accordingly to the following topics:
\begin{itemize}
\item
{\bf Effects of the negative-energy components of the propagators}-
where we compare the QP amplitudes calculated with positive-energy components
of the intermediate propagators ($\eta=0$), with the corresponding 
amplitudes including the full propagators ($\eta=1$). Moreover, by comparing
the amplitudes EH/BbS with the Sp/Th ones, we separate the effects originated
from $I_2 (-\omega_{\footnotesize \mbox{k}}^-)$,
present only in the first pair of equations. We note that accordingly to 
Eqs.\ (\ref{I1}) and (\ref{I2}) the inclusion of the $I_2$ 
residue means extra terms originated by
negative-energy components. In particular, only $I_2$ contains a term involving
negative-energy components of the propagators of both particles.
\item
{\bf Effects of retardation}- where we compare the results of 
Sp/EH equations with and without retardation.
We note that, as shown above, the
Th equation is the instantaneous version of the Sp equation, and that 
BbS equation the instantaneous version of the EH equation (Table I). 
\end{itemize}

In Fig.\ \ref{GrossTh} we show the 
Sp and Th scattering amplitudes calculated with ($\eta=1$) and without
 ($\eta=0$) negative-energy components of the propagators. 
The $+$ label corresponds to
$\eta=0$. As can be seen from the figure, the Sp and
NR amplitudes are indistinguishable at 300 MeV and very close at 1 GeV.
Since the Sp equation includes the combined effects of retardation and
negative-energy state components,
we may conclude that the two types of effects tend 
to add and to approximate the Sp
relativistic calculation to the NR one. We can also see that the 
negative-energy
components decrease the real part of the scattering amplitude for 
Sp and Th formalisms, and,
in the former, retardation enhances this effect.
These conclusions can be drawn for the two energies considered.
We point out that the difference between the 
Sp and NR amplitudes increase with energy as expected.

In Fig.\ \ref{EHBbS} we compare the 
EH and BbS scattering amplitudes, which differ only through the
inclusion of retardation in the former.
It is also shown the ET result.
In all cases the inclusion of negative-energy components 
decreases the real part of the scattering 
amplitude, and approaches the amplitudes to the NR limit. 
More significative is the ET result, 
which has more content of negative-energy components than the EH one,
and is closer to the NR result. We remind here that the ET 
propagator without negative-energy states corresponds to BbS$+$ 
(BbS with only positive-energy states). 
By comparing the retardation effects that are present in EH 
and absent in BbS, we conclude that retardation 
also decreases the real part of the amplitude. 
This is valid for the 
positive-energy versions and for the full versions. 
Once again, we may conclude that the combined effects of retardation and
negative-energy state components tend to approximate the relativistic 
calculations to the NR ones.
The fact that the ET curve, which does not contain retardation, is the closest
to the NR result is explained by the larger contribution of the 
negative-energy
state components in its propagator.

The full scattering amplitude results for the five QP equations 
along with the NR one can be observed in Fig.\ \ref{FigGeral}.
The curves shown confirm the conclusion drawn above, namely that the combined
effect of retardation and negative-energy state contributions tend to 
approximate
the relativistic calculations to the NR ones. 
We point out, however, that the full
result depends upon a delicate interplay of these 
two relativistic effects, and,
in particular, on the amount of the contribution 
of negative-energy state components of
the propagators. This can clearly be seen by comparing the Sp result
(no $I_2(-\omega^-_{\footnotesize \mbox{k}})$) with the EH result
(with $I_2(-\omega^-_{\footnotesize \mbox{k}})$), 
being the first
the closest to the NR limit.

All the previous considerations about the real part of the 
scattering amplitude remain valid for the imaginary 
part with an opposite direction that is, in general retardation 
and negative-energy effects increases the imaginary part 
of the scattering amplitude.

We note that the result from the BbS equation which, accordingly 
to Eqs.\  (\ref{gBbS}) and (\ref{gNR2}), indeed 
generates the NR equation when we replace 
$\frac {m}{E_{\footnotesize \mbox{k}}}$ by 1, 
deviates, the most from the NR result.

\newpage

\begin{figure} 
\centerline{ 
\epsfig{file=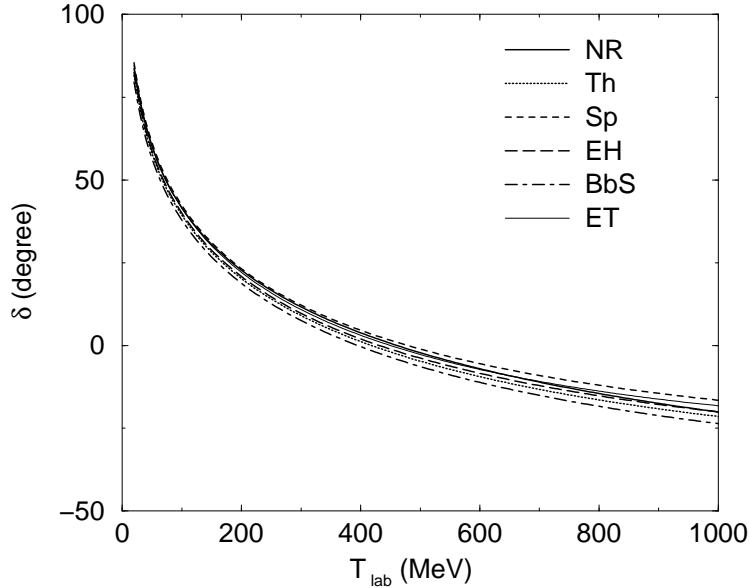,width=8cm,angle=270}}  
\caption{s-wave phase shifts generated with the different QP equations, as
a function of the energy.} 
\label{phases} 
\end{figure}

Finally, as an example, we show in Fig.\ \ref{phases} the phase shifts for
the $l=0$ partial wave, generated by the different QP equations,
as a function of the incoming energy.
As can be seen from the figure, at low energies the
different equations generate similar results, but at higher energies
discrepancies are obvious. In this partial wave the phases generated
by the EH and Th equations are the closest to the NR ones,
contrarily to what happens with the full T-matrix, where the
Sp and the ET are the closest. This fact has a very simple interpretation,
stemming from the fact that at high energies the contribution to the
transition matrix of the more peripherical partial waves 
becomes quite important,
and therefore looking only to the s-wave channel is a very limited analysis. 
This conclusion is also achieved in the work of Ref.\ \cite{Elster98}.

\section{Conclusions} 
\label{secCon} 

We consider a family of relativistic quasipotential scattering equations,
and study the effects of retardation and negative-energy state contributions in
the propagators. We restrict our calculation, for the time being, to
the scalar case. Comparison with the NR equation is provided. The numerical
calculations of the $T$ matrices  were done without partial wave decomposition.
Instead, we solved the two-dimensional integral equation, and
the accuracy of the method was checked through identity provided by the 
optical theorem.
The method implied a discretization of the integral variables with
17-25 grid points for the magnitude of the relative momentum,
and 20-50 grid points for the angular variables. The NR solutions 
were compared
with the results of Ref.\ \cite{Elster98}, where also the partial 
wave decomposition
has been avoided. This comparison provided an extra check to our method.

We conclude that, in general, retardation decreases 
the real part of the scattering amplitude, and increases the imaginary 
part of the same amplitude. 

The inclusion of 
negative-energy state contributions induces a similar effect.

The equations that include the $I_2$ term, i.e. the ones containing 
additional negative-energy state effects, 
deviate the most from the NR results.
We found also that the Sp and the ET equations 
generate the scattering amplitudes
which are the closest to the NR result. 
We also point out that the Sp (with retardation but no $I_2$ terms)
and the ET equations (without retardation but with $I_2$ terms)
consider, in an
effective way, some contribution from the crossed-box diagrams.
On the other hand, the ET equation, on top of the $I_2$ term, includes  
further
negative-energy state effects, associated with its specific way 
of describing effectively
the crossed-box diagrams. These additional terms compensate 
partially the absence
of retardation. We may therefore conclude that the 
inclusion of crossed-box diagrams,
even in an effective way, approaches the QP results to NR ones.

It is interesting to note that the way the Spectator formalism 
deals with negative-energy
state components, retardation and effective crossed-box diagrams, 
makes the corresponding
amplitudes the closest to the NR limit, even at high energies.
Finally we note that the relative uncertainty in the real part of the 
scattering $T$ matrix due to different dynamical equations
depends on the scattering angle 
and is the narrowest near the forward-scattering region.
In the backward-scattering, it can be as large as 40\%.
This could be expected since in this region the momentum 
transfer is larger.   

In this work we restrict ourselves to spin zero particles.
However, calculations to account relativistically for the spin $1/2$, 
such that our formalism can be applied to the 
nucleon-nucleon system, is on progress.
In such a more realistic calculation, the uncertainty 
band found in this work for the $T$ matrix 
will most likely imply differences in the nucleon-nucleon 
interaction, compatible with the scattering data and 
originated by the different dynamical formalisms.

\newpage

\bigskip
\begin{center}
{\bf ACKNOWLEDGMENTS}
\end{center}

The authors wish to thank Franz Gross, J.A. Tjon, and 
D.R. Phillips for very helpful discussions. The work was performed 
under Grant PRAXIS BD/9450/96 from Funda\c c\~ao para 
a Ci\^encia e Tecnologia.
  
\appendix 
 
\section{Solving scattering equation without  
partial wave decomposition} 
\label{NoPartialWave} 
 
In order to solve the 2-dimensional integral Eq.\ (\ref{newQP})  
in the variables k and v, we discretize this variables using  
a gaussian grid with the following properties: 
\begin{itemize} 
\item  
Linear grid of $N_u$ points $u_j$ the variable $v \in \; ]-1,1[$,  
with  weight $h_j$. 
\item 
Linear grid of $N_p$ even points $x_i$ the variable $x \in\;  ]0,1[$,  
with  weight $w_i$. The correspondent momentum points are obtained  
through the transformation 
\ba 
& &\mbox{p}_i=\Lambda \frac{x_1}{1-x_i}, \\ 
& &w_i^\prime=\frac{\Lambda}{(1-x_i)^2} w_i, 
\ea  
where $\Lambda$ is a free parameter.  
We and add the point $x_{N_p+1}=1/2$ with $w_{N_p+1}^\prime=0$. 
\end{itemize} 
 
We take $\Lambda=p$ the on-mass-shell initial momentum.  
With this choice the evaluation of the principal  
part in (\ref{newQP}) is simplified  
\be 
{\cal P} \int_0^\infty  
f(\mbox{k}) \mbox{dk}  \simeq \sum\limits_{i=1}^{N_p} f(\mbox{p}_i) 
w_i^\prime.  
\ee 
The point $\mbox{p}_{N_p+1}=\mbox{p}$ 
corresponding to the pole is not included  
in the sum, hence, the function have no singularities.   
 
We take the outcoming variables $\mbox{p}^\prime$ and $u$ from  
the same grid. In that case the integral equation (\ref{newQP})  
is transformed into a linear algebraic system of equations  
with the unknown $T_{(ij)}=T(\mbox{p}_i,u_j;\mbox{p},1;W)$. The brackets  
in $ij$ indicates that this indices can be contracted in only  
one. The linear system can be written as  
\be 
C \cdot T = V, 
\ee 
where  
\be 
V_{(ij)}=V(\mbox{p}_i,u_j;\mbox{p},1;W). 
\ee 
The matrix $C$ are given by  
\be 
C=I+A+iB, 
\ee 
where $I$ in the unity and  
\ba 
A_{(ij),(i^\prime j^\prime)}&=& 
\frac{\mbox{p}_{i^\prime}}{(2 \pi)^2}  
\frac{f(\mbox{p}_{i^\prime};W)}{E_{p_{i^\prime}}-\frac{W}{2}} 
\bar V(\mbox{p}_i,u_j;\mbox{p}_{i^\prime},u_{j^\prime};W) \\ 
B_{(ij),(i^\prime j^\prime)}&=& 
\frac{1}{4 \pi} \frac{(2m)^2 \mbox{p}}{W}  
\bar V(\mbox{p}_i,u_j;\mbox{p}_{i^\prime},u_{j^\prime};W) 
\delta_{i^\prime,N_p+1}
\ea 
The factor $\delta_{i^\prime,N_p+1}$ gives the elastic-cut  
constraint and fixes the momentum $\mbox{p}_{i^\prime}=\mbox{p}=
\mbox{p}_{N_p+1}$.  
Note that $A$ is zero if $i^\prime=N_p+1$, and $B$  
is always zero except for  $i^\prime=N_p+1$.  
Accordingly, the matrix elements of $A+iB$ are purely  
real or imaginary.

\end{document}